\documentclass[sigconf]{acmart}

\AtBeginDocument{%
  \providecommand\BibTeX{{%
    \normalfont B\kern-0.5em{\scshape i\kern-0.25em b}\kern-0.8em\TeX}}}

\copyrightyear{2024}
\acmYear{2024}
\setcopyright{acmlicensed}\acmConference[WWW '24 Companion]{Companion Proceedings of the ACM Web Conference 2024}{May 13--17, 2024}{Singapore, Singapore}
\acmBooktitle{Companion Proceedings of the ACM Web Conference 2024 (WWW '24 Companion), May 13--17, 2024, Singapore, Singapore}
\acmDOI{10.1145/3589335.3651478}
\acmISBN{979-8-4007-0172-6/24/05}

\usepackage{amsmath,amsfonts}
\usepackage[linesnumbered,ruled,vlined]{algorithm2e}
\usepackage{pifont}
\usepackage{amsthm}
\usepackage{algpseudocode}
\usepackage{tcolorbox}
\usepackage{array}
\usepackage[caption=false,font=normalsize,labelfont=sf,textfont=sf]{subfig}
\usepackage{textcomp}
\usepackage{stfloats}
\usepackage{url}
\usepackage{verbatim}
\usepackage{graphicx}
\usepackage{ulem}
\usepackage{amsfonts}
\usepackage[utf8]{inputenc}
\usepackage{setspace}
\usepackage{cleveref}
\usepackage[utf8]{inputenc}
\crefname{section}{§}{§§}
\Crefname{section}{§}{§§}
\usepackage{caption}
\usepackage{booktabs}
\usepackage{makecell}
\usepackage{MnSymbol}
\usepackage{utfsym}
\usepackage{filecontents, pgffor}
\usepackage{balance}
\usepackage{latexsym}
\usepackage{color}
\usepackage{listings, xcolor}
\definecolor{verylightgray}{rgb}{.97,.97,.97}

\lstdefinelanguage{Solidity}{
	keywords=[1]{anonymous, assembly, assert, balance, break, call, callcode, case, catch, class, constant, continue, constructor, contract, debugger, default, delegatecall, delete, do, else, emit, event, experimental, export, external, false, finally, for, function, gas, if, implements, import, in, indexed, instanceof, interface, internal, is, length, library, log0, log1, log2, log3, log4, memory, modifier, new, payable, pragma, private, protected, public, pure, push, require, return, returns, revert, selfdestruct, send, solidity, storage, struct, suicide, super, switch, then, this, throw, transfer, true, try, typeof, using, value, view, while, with, addmod, ecrecover, keccak256, mulmod, ripemd160, sha256, sha3}, % generic keywords including crypto operations
	keywordstyle=[1]\color{blue}\bfseries,
	keywords=[2]{address, bool, byte, bytes, bytes1, bytes2, bytes3, bytes4, bytes5, bytes6, bytes7, bytes8, bytes9, bytes10, bytes11, bytes12, bytes13, bytes14, bytes15, bytes16, bytes17, bytes18, bytes19, bytes20, bytes21, bytes22, bytes23, bytes24, bytes25, bytes26, bytes27, bytes28, bytes29, bytes30, bytes31, bytes32, enum, int, int8, int16, int24, int32, int40, int48, int56, int64, int72, int80, int88, int96, int104, int112, int120, int128, int136, int144, int152, int160, int168, int176, int184, int192, int200, int208, int216, int224, int232, int240, int248, int256, mapping, string, uint, uint8, uint16, uint24, uint32, uint40, uint48, uint56, uint64, uint72, uint80, uint88, uint96, uint104, uint112, uint120, uint128, uint136, uint144, uint152, uint160, uint168, uint176, uint184, uint192, uint200, uint208, uint216, uint224, uint232, uint240, uint248, uint256, var, void, ether, finney, szabo, wei, days, hours, minutes, seconds, weeks, years},	% types; money and time units
	keywordstyle=[2]\color{teal}\bfseries,
	keywords=[3]{block, blockhash, coinbase, difficulty, gaslimit, number, timestamp, msg, data, gas, sender, sig, value, now, tx, gasprice, origin},	% environment variables
	keywordstyle=[3]\color{violet}\bfseries,
	identifierstyle=\color{black},
	sensitive=true,
	comment=[l]{//},
	morecomment=[s]{/*}{*/},
	commentstyle=\color{gray}\ttfamily,
	stringstyle=\color{red}\ttfamily,
	morestring=[b]',
	morestring=[b]"
}

\begin{document}
\title{Characterizing the Solana NFT Ecosystem}

\author{Dechao Kong}
\email{3037317181kdc@gmail.com}
\affiliation{%
  \institution{School of Cyberspace Security}
  \institution{Hainan University}
  \city{Haikou}
  \country{China}
}

\author{Xiaoqi Li}
\authornote{The corresponding author}
\email{csxqli@gmail.com}
\orcid{1234-5678-9012}
\affiliation{%
  \institution{School of Cyberspace Security}
  \institution{Hainan University}
  \streetaddress{P.O. Box 1212}
  \city{Haikou}
  \country{China}
}

\author{Wenkai Li}
\email{liwenkai871@gmail.com}
\affiliation{%
  \institution{School of Cyberspace Security}
  \institution{Hainan University}
  \city{Haikou}
  \country{China}
}

% \date{February 2023}

% \renewcommand{\shortauthors}{Wenkai and Xiaoqi, et al.}

\begin{abstract}
  Non-Fungible Tokens (NFTs) are digital assets recorded on the blockchain, providing cryptographic proof of ownership over digital or physical items. Although Solana has only begun to gain popularity in recent years, its NFT market has seen substantial transaction volumes. In this paper, we conduct the first systematic research on the characteristics of Solana NFTs from two perspectives: longitudinal measurement and wash trading security audit. We gathered 132,736 Solana NFT from Solscan and analyzed the sales data within these collections. Investigating users' economic activity and NFT owner information reveals that the top users in Solana NFT are skewed toward a higher distribution of purchases. Subsequently, we employ the Local Outlier Factor algorithm to conduct a wash trading audit on 2,175 popular Solana NFTs. We discovered that 138 NFT pools are involved in wash trading, with 8 of these NFTs having a wash trading rate exceeding 50\%. Fortunately, none of these NFTs have been entirely washed out.
\end{abstract}




\maketitle

\section{INTRODUCTION}
Solana NFTs are digital assets established on the Solana blockchain, representing unique projects or content segments under authorized ownership. Compared to the token use of Bitcoin or Ethereum, NFT favors unique and irreplaceable digital assets, and NFT has practical applications in scenarios such as gaming, art, and virtual real estate. Compared to other blockchains, Solana stands out with its superior transaction processing capacity and extremely low transaction fees. It can handle thousands of transactions per second, leading to a massive user base for Solana. These features attract the attention of NFT artists and crypto users. With Solana's distinctive features, Solana NFTs have entered the public eye since 2021 and have since maintained tens of millions of dollars in sales volume each month\cite{cryptoslam}. The global number of cryptocurrency users has surpassed 300 million \cite{crypto}, and this large user base has contributed to the increased circulation of cryptocurrencies. The growing user community is driving the prosperity of Solana NFTs.

The inherent security of NFTs\cite{das2022understanding,wang2023nfts}, coupled with the lack of trading regulation exposes those who participate in NFT investments to potential risks. The price fluctuations of NFTs are easily influenced by external commentary. The issues of fraud in the process from NFT minting to user asset allocation\cite{pungila2022new}, user information theft \cite{pianzi}, account asset security\cite{li2021clue,li2020characterizing}, user privacy disclosure\cite{li2305overview}, and NFT asset storage security further exacerbate prejudices towards this field. Utilizing the robot operating system\cite{zhang2022authros} for NFT asset data sharing reduces errors and ensures data security. The burden of token verification falling on the users greatly facilitates NFT network scams. For example, hackers lured traders to buy fake Banksy NFT for \$336,000.

The related work on Solana NFTs currently extends beyond merely focusing on the sales of NFTs. It also explores practical applications. Artha et al.\cite{artha2022implementation} combined smart contracts and Solana NFT to convert certificates to NFTs to solve the certificate packing problem. Zocca et al.\cite{zocca2023combining} designed and implemented a system to counter the phenomenon of counterfeit luxury goods. using NFT and NFC tags. Solana NFTs are deployed to the blockchain through smart contracts. Presently, LLMs are employed for the detection of these smart contracts\cite{mao2024automated}, enhancing the security of the NFTs' underlying assets. DeFi\cite{li2022survey} and NFTs, as emerging financial systems, still have issues in the realm of security. Particularly concerning is market manipulation in NFT sales, where malicious actors exploit wash trading to artificially inflate prices and profit from legitimate transactions. Current methods for detecting Solana NFT wash trading primarily rely on Strongly Connected Components (SCC)\cite{wen2023nftdisk} in graphs for transaction analysis. However, users can easily create multiple trading accounts, making it difficult to address wash trading patterns characterized by high-frequency, low-value transactions.

In this paper, we conduct a systematic analysis of Solana NFT by analyzing its sales data. We begin by extracting Solana NFT collections from Solscan, gathering NFT sales data based on the officially provided API, and finding that the top users in the NFT occupy a significant share of the market and can influence the transactions in the NFT market. Then, utilizing the Local Outlier Factor\cite{breunig2000lof} algorithm to identify wash trading nodes in the Solana NFT transaction, calculate clusters and outliers within the NFT transaction node dataset to flag suspicious NFT. Experimental evaluation of wash trading detection on 2,175 popular NFTs. We discover that 138 NFT collections have wash trade behavior, revealing the NFT involved in fraudulent trades from the security aspect. The main contributions of this paper are as follows:

\begin{itemize}
\item To the best of our knowledge, we represent the first systematic longitudinal analysis of Solana NFTs to elucidate its market evolution and malicious behaviors.
\item We employ the local outlier factor algorithm to detect wash trading in Solana NFTs. By calculating the local deviation values of NFT transaction nodes adjacent to other nodes, we identify wash trading nodes.
\item We open-source related codes and data at \url{https://figshare.com/s/c3cd64790e13c377cf84}.
\end{itemize}

\section{METHODOLOGY}
\subsection{Datasets}
We mine NFT sales data from Solscan \cite{solscan}, which is a Solana blockchain explorer that allows for online tracking of Solana NFT activity. Statistical data shows that the cumulative transaction volume of Solana NFTs has exceeded 2 billion US dollars, with more than 1.6 million traders \cite{solana}. We utilize the Solscan blockchain explorer, using the official API to build a dataset. To make the data easier to analyze and manage, we perform queries within a specific range using the official API and set a specific size of sliding window to obtain the sales data recorded in the Solscan blockchain explorer from January 2022 to October 2023. In total, we obtain 132,736 different series of NFTs and 28,706,698 sales data on the Solana blockchain.

\subsection{Data Overview}
Table \ref{tab:freq1} lists the sales of Solana NFTs by year. The transaction volume in each quarter of 2022 was higher than that of 2023. The sales volume decreased by 69\% from January 2022 to October 2023. The transaction volume began to decline in the first and second quarters of 2022, and the SOL token price is falling at this time, suggesting that Solana (SOL) price volatility is closely tied to Solana NFT sales.

In our analysis of the top NFT holder accounts, we identify a significant number of accounts labeled as Pro NFT Minter or Pro NFT Holder. These accounts are consistently active traders with a high average daily transaction volume in their wallets. The holdings of NFT holders affect the overall value of the NFT. Table \ref{tab:freq2} shows the ownership information of the top holder with a rank of 1 in each NFT. The proportion of holdings by the top holder in the NFT collection is divided into four sections: 0\%-25\%, 25\%-50\%, 50\%-75\%, and 75\%-100\%. Among them, holders who own between 75\%-100\% have an average of 152.36 NFTs, with some users even owning more NFTs. In the second quarter of 2022, the account with the most NFTs in the top holder's collection had as many as 127,612 NFTs. However, owning too many NFTs from this collection did not increase the economic value of this NFT, proving that monopolistic behavior does not increase the value of the NFT.

\begin{table}[h]

  \caption{NFT Quarterly Sales Volume Analysis}
  \label{tab:freq1}
  \begin{tabular}{ccc}
    \toprule
    Quarter &2022 & 2023\\
    \midrule
    \texttt Q1 & \$823,148,455.54 & \$340,000,906.42 \\
    \texttt Q2 & \$777,484,916.62 & \$221,347,550.62\\
    \texttt Q3 & \$250,890,363.26& \$97,285,641.64\\
    \texttt Q4 & \$252,846,399.38& - \\
    \hline
    \texttt Total & \$2,104,370,134.80 & \$658,634,098.68
    %\bottomrule
  \end{tabular}
  \vspace{-2ex}
\end{table}

\begin{table}[H]
  %\fontsize{7.6}{14}\selectfont  %{字体尺寸}{行距}
  \caption{The Top Holder with a Rank of 1}
  \label{tab:freq2}
  %%\vspace{2ex}
  \begin{tabular}{ccc}
    \toprule
    Interval &Percentage & Average NFT Ownership\\
    \midrule
    \texttt 0\%-25\% & \ 37.76\% & \ 57.95 \\
    \texttt 25\%-50\% & \ 24.04\% & \ 70.03\\
    \texttt 50\%-75\% & \ 27.62\% & \ 52.8 \\
    \texttt 75\%-100\% & \ 10.58\% & 152.36 \\
    \bottomrule
  \end{tabular}
  \vspace{-2ex}
\end{table}

\begin{figure}[H]
  \centering
  \includegraphics[width=\linewidth, height=5cm]{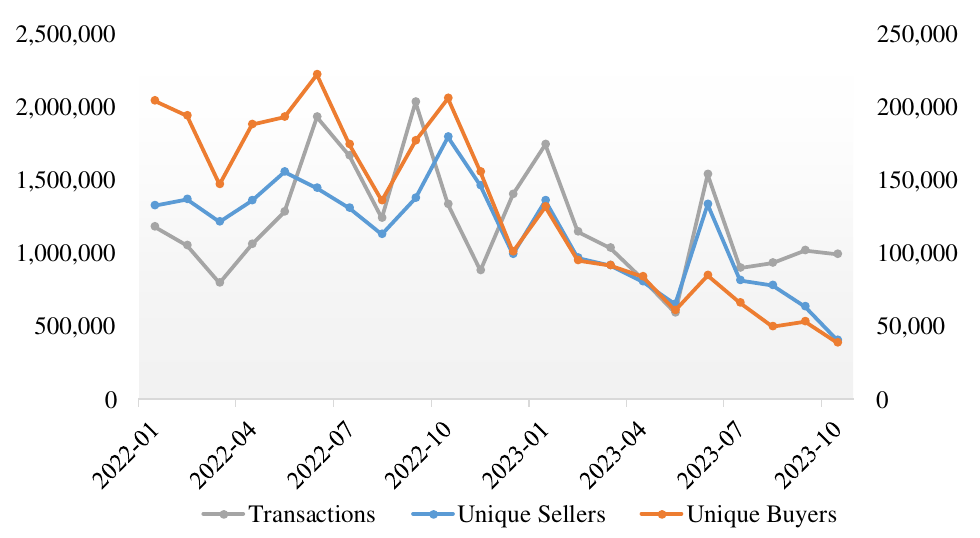}
  \caption{Timeline of Buyers, Sellers, and Transaction Volume. Transactions serve as the primary axis on the left, while the number of buyers and sellers constitutes the secondary axis on the right.}
  %% \Description{A woman and a girl in white dresses sit in an open car.}
  \label{fig:1}
  \vspace{-2ex}
\end{figure}

\begin{figure}[H]
  
  \centering
  \includegraphics[width=\linewidth, height=5cm]{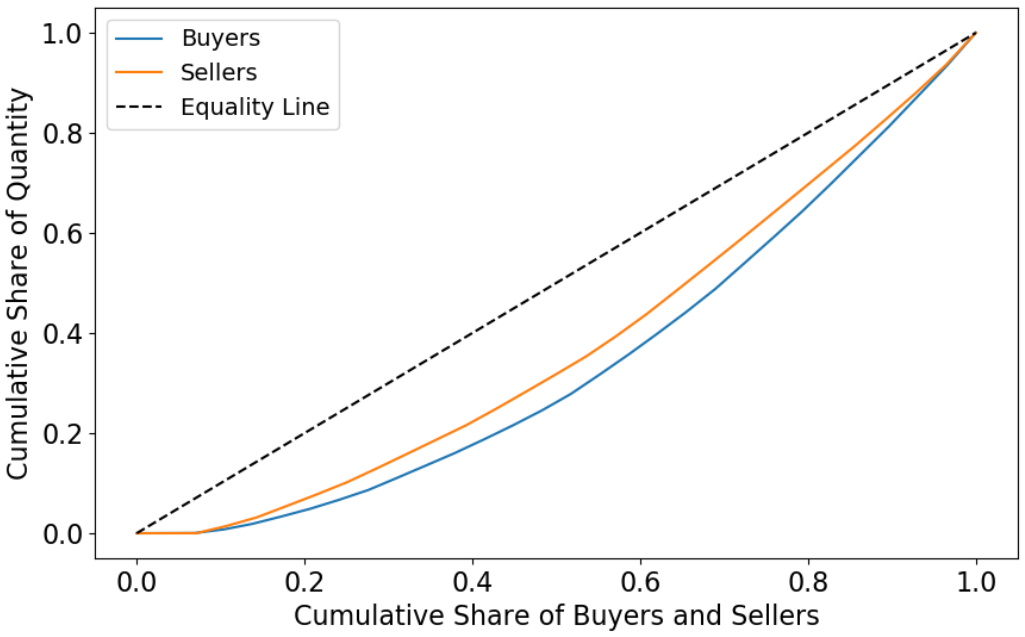}
  \caption{Lorenz Curves Representing Cumulative Purchase Quantity for Buyer and Seller Percentiles}
  %% \Description{A woman and a girl in white dresses sit in an open car.}
  \label{fig:2}
  \vspace{-2ex}
\end{figure}

Figure \ref{fig:1} shows the relationship between buyers, sellers, and transaction volumes by year. Transaction volume began to rise in the year Solana NFT appeared in 2022, with the first peak in June. User volume also peaked in this month, and transactions peaked in September 2022, but the number of users year-on-year was not as high as in June. 2023 also saw two peaks, but there was a general downward gradient in both transaction volume and user volume compared to 2022. This shows that the growth and decline in trading volume are closely related to the number of users.

Figure \ref{fig:2} shows the cumulative percentage of purchases by buyers and sellers, indicating that the distribution of purchases by buyers or sellers tends toward higher percentiles. This is a typical feature of trading in the NFT market. 60\% of market transactions are controlled by 40\% of users, but this is not as extreme as the sales situation in the OpenSea trading market\cite{white2022characterizing}.

\section{MARKET MANIPULATION}
In this section, we will delve into the current state of market manipulation, with wash trading being the focal point. Due to the characteristics of Solana's blockchain technology, traders can engage in NFT transactions without the need to disclose their real identities. Compared to traditional assets, this feature makes NFTs more appealing for wash trading. With the substantial transaction volume of Solana NFTs, it becomes even easier to profit from such wash trading activities.

\subsection{Wash Trade Detection}
Our work employs Local Outlier Factor detection to identify wash trading in NFT transactions, with a particular focus on uncovering wash trading activities within the NFT market.

\noindent \textbf{Local Outlier Factor: }The transaction addresses are generalized as nodes and anomalous data nodes are identified by detecting the local deviation of a given node relative to its neighboring nodes. The LOF value is used to ascertain whether a node is an outlier and the extent of its outlier status. In the context of NFT transactions, there are specific patterns of wash trading, such as frequent small transactions, irregular transaction times, and transaction amounts higher than the average selling price. Normal transactions tend to form high-density areas with low degrees of outlier status, thus the LOF algorithm can be effectively used to detect anomalous nodes in NFT transactions.

\begin{equation}
{C_i=\frac{\left | \left \{ e_{jk} : v_j,v_k\in N_i,e_{jk}\in E \right \}  \right | }{k_i(k_i-1)}}
\end{equation}

\begin{equation}
lrd_{k}(y_{i})=\frac{\left | N_{k}(y_{i})  \right | }{ {\textstyle \sum_{y_{ip}\in N_{k}(y_{i})}^{}RD_{k}(y_{i},y_{ip})} }
\end{equation}

\begin{equation}
LOF_{k}(y_{i})=\frac{ {\textstyle \sum_{y_{ip}\in N_{k}(y_{i})}^{}}lrd_{k}(y_{ip}) }{\left | N_{k}(y_{i}) \right |*lrd_{k}(y_{i}) } 
\end{equation}

\noindent \textbf{Node clustering coefficient: }The clustering coefficient of a node greatly influences its outlier factor value, so it is necessary to calculate the clustering coefficient between various nodes before identifying local anomalous nodes\cite{yin2023network}. The clustering coefficient of a node set is calculated using equation 1, where ${e_{jk}}$ represents an edge connecting nodes ${v_j}$ and ${v_k}$, ${N_i}$ is the set of neighbor nodes of node ${v_i}$, and ${k_i}$ is the number of neighbor nodes of node i.

\noindent \textbf{Node local outlier factor}: The local reachability density of a sample point $y_{i}$ can be calculated using equation 2. Treat nodes as samples. Assuming there are two sample points, $y_{i}$ and $y_{j}$, d($y_{i}$,$y_{j}$) represents the distance between these two sample points, $d_{k}$($y_{i}$)=d($y_{i}$,$y_{ik}$) represents the K-distance neighbor distance of data point $y_{i}$, and the symbol $N_{k}$($y_{i}$) represents the K-distance neighborhood set of $y_{i}$. The reachable distance between nodes$y_{i}$ and $y_{j}$ can be calculated as $RD_{k}(y_{i},y_{j})$=max{$d_{k}(y_{i},d(y_{i},y_{j}))$}. The local outlier factor of sample point $y_{i}$, $LOF_{k}(y_{i})$, is defined as in equation 3. From the above equations, it can be seen that if the LOF value of sample point $y_{i}$ is relatively large, it indicates that its local density is small and the degree of deviation from other sample points in the neighborhood is higher. Such a point can be considered a suspicious node.

\subsection{Experiments and Result}
We perform the acquisition and experimentation of the dataset on a host device (Intel(R) Core(TM) I7-12700 CPU + 32GB RAM). The k-distance neighbor parameter in LOF is set to 20.

Due to the large number of NFTs gathered, we selected 2,492 popular NFTs with varying degrees of trading volume for analysis. These collections span across 13 NFTMs and encompass 69,736 NFT transactions, with a total transaction amount of \$5,340,344. We used these popular NFTs as our dataset and applied the LOF detection model to the NFT transaction sets for NFT wash trade detection.

\begin{figure}[H]
     \centering
    \includegraphics[height=5cm]{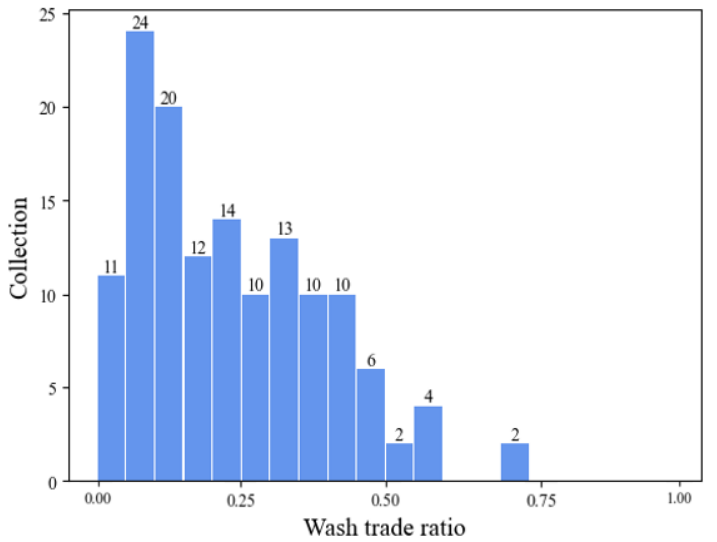}
    \caption{Distribution of Wash Trading Rates among NFT Collections}
    \label{fig:3}
    \vspace{-2ex}
\end{figure}

We detected 2,175 suspicious nodes associated with 46,612 transactions involving an amount of \$3,744,092. Suspicious nodes account for approximately 12.3\%. Most wash trades fall into the category of high-frequency, low-value transactions \cite{kang2022market}, with a few nodes capable of generating multiple wash trades. This also explains the situation where the proportion of suspicious nodes is low while the number of wash trades is high. Furthermore, among the 2,492 NFTs in our dataset, 13 collections have a transaction volume exceeding \$50,000. Additionally, 138 NFTs show varying degrees of evidence of wash trading.

We define the Wash Trading Ratio (WTR) of a collection as the ratio of the total transaction volume involving wash trades. In other words, when WTR is 1, all transactions in the NFT are suspected of wash trading. In Figure \ref{fig:3}, we show the distribution of the WTR for collections suspected of wash trading. Among all NFTs with wash trades, 81 (58.69\%) NFTs have less than 25\% wash trades (WTR < 0.25), and only 5.79\% of collections have wash trades exceeding half of the total transactions. Fortunately, no NFTs were completely misused.

\begin{figure}[H]
    \centering
    \includegraphics[ height=5cm]{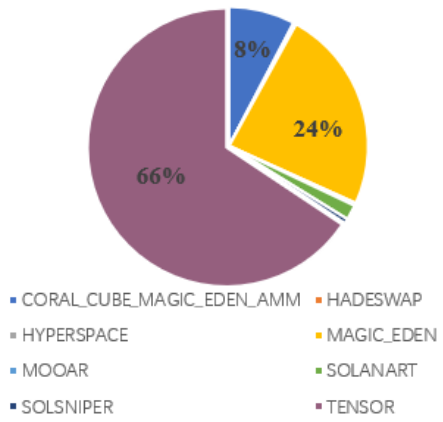}
    \caption{The Comparative WTR across Different Marketplaces}
    \label{fig:4}
    \vspace{-2ex}
\end{figure}

Figure \ref{fig:4} illustrates the proportion of wash trades occurring across various NFTMs. In the TENSOR\cite{tensor} NFT marketplace, the volume of wash trading significantly exceeds that of the other seven NFTMs, accounting for as much as 66\%. As the marketplace with the largest transaction volume for Solana NFTs, TENSOR has a noticeably higher frequency of wash trades compared to other NFTMs. MAGIC\_EDEN \cite{magicende}, as the second largest marketplace for Solana NFTs, also has a high wash trade rate of 24\%. Combined, these two marketplaces account for 90\% of the total volume of wash trades.

There are 8 collections with a wash trading rate above 50\%, each containing NFTs that have been subjected to varying degrees of wash trading, as shown in Table \ref{tab:freq3}. The NFT collection Froganas has 5,555 NFTs, out of 195 NFTs with transaction records, 114 were exploited to inflate the trade volume of NFTs, involving price manipulation totaling \$76,852.51. Following this, the trade volume and price of Froganas increased nearly tenfold.

\begin{table}[H]
  \setlength{\abovedisplayskip}{0cm} %%% 3pt 可自行设置
  \setlength{\belowdisplayskip}{0cm}
  \caption{Collections with WTR Exceeding 50\%}
  \label{tab:freq3}
  \vspace{-2ex}
  \begin{tabular}{cccc}

    \toprule
    Collection&Tokens&Wash tokens&Wash rate\\
    \midrule
    \texttt Solana Maxi & 125 & 70 & 56.00\% \\
    \texttt Node Whales & 115& 61  & 53.04\% \\
    \texttt Gosty & 41 & 23 & 56.09\% \\
    \texttt frank & 101 & 56& 55.44\% \\
    \texttt Froganas & 195 & 114 &58.46\% \\
    \texttt Soul Again & 55 & 40 &72.72\% \\
    \texttt Gaimin Gladiators & 65 & 35 &53.84\% \\
    \texttt SOLANA FROG & 85 & 62 &72.94\% \\
    \bottomrule
  \end{tabular}
\end{table}

\section{CONCLUSION}
In this paper, we obtain 132,736 Solana NFT collections and 28,706,698 sales data from the Solana blockchain. We examine the relationship between users and transaction volume, demonstrating that top holders can influence the majority of transactions within an NFT. After conducting the distribution of purchases, we find that Solana NFT allows a small segment of the population to dominate. Furthermore, we employ the LOF algorithm to analyze wash trading in popular NFT collections, aiming to alert users investing in NFTs. In the future, we will also analyze the security of Solana NFTs at different stages, from minting to assets.

\normalem
\bibliographystyle{ACM-Reference-Format}

\bibliography{main}

\end{document}